\documentclass{ws-p8-50x6-00}

\begin{document}

\title{\lowercase{$p$--}shell nuclei and two-frequency shell model 
with a realistic effective interaction}

\author{A. Gargano, L. Coraggio, A. Covello and N. Itaco}

\address{Dipartimento di Scienze Fisiche, Universit\`{a} di Napoli 
Federico II\\
and Istituto Nazionale di Fisica Nucleare\\
Complesso Universitario di Monte S. Angelo, Via Cintia, 80126 Napoli, Italy}

\author{T. T. S. Kuo}

\address{Department of Physics, SUNY, Stony Brook, New York 11794,USA}


\maketitle

\abstracts{
We have studied $p$-shell nuclei using a two-frequency shell-model approach with an effective 
interaction derived from the Bonn-A nucleon-nucleon potential by means of a $G$-matrix 
folded-diagram method. First, we briefly describe our derivation of the effective interaction in 
a model space composed of harmonic oscillator wave functions with two different length
parameters, $b_{\rm in}$ and $b_{\rm out}$, for the core and the valence orbits, respectively.
Then we present some selected results of our calculations. We show that a good
agreement with experiment is obtained, which is definitely better than that provided by
a standard one-frequency calculation. A comparison with results obtained from
large-basis shell-model calculations is also made.}     

\section{Introduction}

The $p$-shell nuclei have long been the subject of several theoretical
investigations. In recent years, however, special interest in the study of
these nuclei has been motivated by new relevant information on
their structure. One of the main results has been the observation for some
neutron-rich nuclei, such as $^6$He and $^{11}$Li, of abnormally large
interaction and reaction cross-sections.\cite{tan1,tan2} These nuclei have a very small
one- and two-neutron separation energy and have been described as having a
halo structure,\cite{han87} namely an extended neutron distribution surrounding a
tightly bound inner core. Generally, it has been observed that the $p$-shell
nuclei have a rather extended density profile.

Within the framework of the standard shell-model, the description of these
nuclei requires the use of multi-shell model spaces.$^{4-6}$ In particular,
large-basis no-core shell-model calculations have been performed \cite{nav98} making use
of effective interactions derived from modern nucleon-nucleon
($NN$) potentials.
The use of very large model spaces, however, makes in general the
calculations quite complicated. 

Recently, a two-frequency shell-model (TFSM)
approach with a realistic $G$-matrix effective interaction has been
developed \cite{kuo96tf,kuo97tf} to treat light nuclei near the drip lines.
In this approach, the calculation of the effective interaction involves use
of harmonic oscillator wave functions of two different oscillator constants
$\hbar \omega_{\rm in}$ and $\hbar \omega_{\rm out}$ for the core and
valence orbits, respectively (the length parameters for
these orbits will be denoted by $b_{\rm in}$ and $b_{\rm out}$, $b=(\hbar/m\omega)^{1/2}$).
This provides a simpler treatment of this kind of nuclei in that use
of very large active spaces may be avoided.
In fact, the effective
interaction derived in the above way from the realistic $NN$ potential should take into account
the effects of correlations  which are not explicitly included in
the model space.
We shall see in Sec. 3 that, in spite of its simplicity, the TFSM approach
yields results which are quite satisfactory. Actually, it turns out that
they are comparable to, or even better than, those obtained from large-basis
shell-model calculations.

\section{Two-Frequency Shell Model}

In the TFSM approach the effective interaction $V_{\rm eff}$ is derived from
the free $NN$ potential by way of a $G$-matrix folded
diagram method. This method has been developed within the framework of
the ordinary (one-frequency) shell model and its description
including a complete list of references can be found in Refs. 9 and 10. 
Here, we only give the essentials of the derivation of $V_{\rm eff}$
focusing attention on those features which are peculiar to the
TFSM. 

The first step for the derivation of  $V_{\rm eff}$ is the choice of a closed
core and the single-particle (sp) orbits over which the valence nucleons
are distributed, that is the model space (usually referred to as $P$ space)
within  which the effective interaction is derived.
It should be noted here that $V_{\rm eff}$ is usually derived 
for two-valence particles or holes outside a doubly closed core.
For instance, the $P$ space for $p$-shell nuclei consists of a closed
$(0s_{1/2})^{4}$ core  with the two-valence nucleons confined in the 
$0p_{3/2}$ and $0p_{1/2}$
orbits. As usual, the $P$ space is defined in terms of the 
eigenvectors of the harmonic oscillator. Within the TFSM 
approach, sp wave functions of two different 
length parameters $b_{\rm in}$  and $b_{\rm out}$ are used, the former for the 
inner core orbits and the latter for the outer valence orbits. Actually,
as we shall see below,  also
orbits above the $p$ shell come into play  as intermediate states
in the evaluation of the effective interaction. In this context,
it should be noted that two sp wave functions with the same $l$ and $j$ 
values but different length parameters are in general not orthogonal. 
To assure the orthonormality condition for these states, in the present 
calculations we use  $b_{\rm in}$ also for all 
outer orbits with $l=0$.   

At this point, $V_{\rm eff}$ could be derived for the chosen model space
by means of a perturbation calculation in terms of the $NN$ potential.
However, the strong repulsive core contained in all realistic $NN$
potential makes a perturbative  treatment  meaningless. 
This problem is overcome by introducing the reaction
matrix $G$. This matrix is defined by the integral equation \cite{krench76}

\begin{equation}
G(\omega)=V+VQ_2 \frac{1}{\omega-Q_2TQ_2}Q_2G(\omega),
\end{equation}

\noindent
which can be solved in an essentially exact way by the matrix inversion
method of Tsai-Kuo.\cite{tsai72}
In Eq. (1) $V$ represents the $NN$ potential, $T$ is the two-nucleon kinetic
energy, and $\omega$, commonly referred to as starting-energy, is 
the unperturbed energy of the interacting nucleons.
The operator $Q_{2}$ is a two-body Pauli exclusion operator,
whose complement $P_{2}=(1-Q_{2})$  defines the space within which the $G$
matrix is calculated. The role of $Q_{2}$ in Eq. (1) is to make sure that
the intermediate states of $G$ be outside this space. Note that 
plane-wave functions are employed for the intermediate states while 
$Q_{2}$ is defined in terms of harmonic oscillator wave functions as

\begin{equation}
Q_{2}= \sum_{{\rm all} \ a b} Q(ab) |ab \rangle \langle ab|,
\end{equation}

\noindent
where $Q(ab)=0$, if $b \leq n_{1}$, $a \leq n_{3}$, or $b \leq n_{2}$,
$a \leq n_{2}$, or $b \leq n_{3}$, $a \leq n_{1}$, and $Q(ab)=1$
otherwise. 
The boundary of  $Q_{2}$ is specified
by the three numbers $n_{1}$,
$n_{2}$, and $n_{3}$, each representing a single-particle orbital
(the orbits are numbered starting from the bottom of the oscillator well).
In particular, $n_{1}$ is the number of orbitals below the
Fermi surface of the doubly magic core, $n_{2}$  fixes the orbital
above which  the passive sp states start, and $n_{3}$ denotes 
the limit of the $P_{2}$ space. 

It should be noted that in the calculation of $G$ the space of active
sp states,
i.e. the number of levels
between $n_{1}$ and $n_{2}$, may be different from the model space within 
which $V_{\rm eff}$ is defined. Several arguments for choosing 
the former larger
than the latter are given in Ref. 11. Generally, $n_{2}$ is fixed 
so as to include two major shell above the Fermi surface. In this paper
we consider the $p$-shell nuclei with $^{4}$He treated as core, thus we have
$n_{1}=1$. Then we take $n_{2}=6$ so as to include all the five orbits of 
the $p$ and $sd$ shells above the Fermi surface. As regards the value of
$n_{3}$, it should be infinite, but in practice it is chosen to be 
a large but finite number. Namely, calculations are performed for 
increasing values of $n_{3}$ until numerical results become stable.
For the present case we have found that a choice of $n_{3}=21$ turns
out to be adequate. 

From the above it is clear that the reaction matrix $G$, which depends on 
the space $P_{2}$, is different
for different choices of this space. As already discussed in the Introduction,
in the TFSM the choice of two different length parameters, for the
core and valence wave functions,  seems to be appropriate to describe 
light nuclei such as those of the $p$ shell. It poses, however, some technical
difficulties in the calculation of the $G$ matrix. In fact, transformations 
from two-particle states in the c.m. coordinates to those in the laboratory
coordinates are not as easy to perform as in the case of a unique 
oscillator frequency. We have adopted an expansion procedure \cite{kuo96tf,kuo97tf} to surmount
this problem. Namely, we expand the wave functions with $b_{\rm in}$ in 
terms of
those with  $b_{\rm out}$, or vice versa.

Finally, using the above $G$ matrix, we can now calculate  $V_{\rm
eff}$ in the chosen model space. 
This  interaction, which is energy independent,  can be written 
schematically in operator form as \cite{kuo80}

\begin{equation}
V_{\rm eff} = \hat{Q} - \hat{Q'} \int \hat{Q} + \hat{Q'} \int \hat{Q} \int
\hat{Q} - \hat{Q'} \int \hat{Q} \int \hat{Q} \int \hat{Q} + ~...~~,
\end{equation}

\noindent
where $\hat{Q}$ (referred to as $\hat{Q}$-box) is a vertex function composed 
of irreducible  linked diagrams in $G$, and the integral sign represents a
generalized folding operation .
$\hat{Q'}$ is obtained from $\hat{Q}$ by removing terms of first order in the
reaction matrix $G$. 
After the $\hat{Q}$-box is calculated, $V_{\rm eff}$ is
obtained by summing up the folded-diagram series (3) to all orders by 
means of the Lee-Suzuki iteration method.\cite{suzuki80}
This last step can be performed in an essentially exact way for a given
$\hat{Q}$-box.

As regards the calculation of  the $\hat{Q}$-box, we need to make 
certain approximations, namely  we include only 
diagrams through second order in  $G$. They are precisely the
seven first- and second-order diagrams considered by
Shurpin {\em et al.}\cite{shurp83}
Higher-order diagrams might be important in some cases. For example, in Refs.
16 and 17 the role of third-order diagrams has been investigated within the 
framework of standard shell-model calculations. It was shown there that 
for the $sd$ nuclei
the third-order contributions lead to a change of about $10-15\%$ in 
the effective 
interaction, while they reduce to only $5\%$ or less  
for heavier nuclei (in this case only for the $T=1$ matrix elements). 
In the TFSM approach one may 
expect these higher-order diagrams  are even 
less important. In fact, the magnitude of second order core-polarization
diagrams, representing the most significant corrections to the $G$ matrix, 
become rather small  when the length parameter $b_{\rm out}$ becomes 
significantly
larger than  $b_{\rm in}$.\cite{kuo97tf} This is a consequence of the fact that
when increasing $b_{\rm out}$ we are increasing the average distance
between the core and valence nucleons, thus reducing the overlaps
between them.

\section{Results for $p$-Shell Nuclei}

We present here some results of our calculations for the $p$-shell nuclei. 
They have been obtained by using the OXBASH shell model code.\cite{oxb}

As already mentioned in the preceding section, we have assumed that the
doubly magic $^{4}$He is a closed core and let the valence particles 
occupy the two orbits $0p_{3/2}$ and $0p_{1/2}$. The spacing between these
two levels, $\epsilon_{1/2}-\epsilon_{3/2}$, was fixed at 4.0 MeV,
which corresponds to the excitation energy \cite{nndc} of the first $\frac{1}{2}^{-}$
state in $^{5}$He.  This state, however, is a very broad resonance with a 
large error bar of about $\pm 1$ MeV. For the sp energy 
$\epsilon_{3/2}$ we have taken the experimental one-neutron separation
energy for $^{5}$He, 0.886 MeV.\cite{audi95}
The effective interaction was derived 
from the Bonn-A free $NN$ potential \cite{mach89} following the procedure described
in Sec. 2.

As regards the two length parameters $b_{\rm in}$ and $b_{\rm out}$, 
the former was fixed~\cite{kuo96tf} at 1.45 fm  while the latter was allowed to vary
from 1.45 to 2.50 fm.
To fix the value of $b_{\rm out}$, we have calculated the ground-state
energies for several nuclei with $6 \leq A \leq 8$. The corresponding 
binding energies have been then obtained by making use of  the experimental 
ground-state
binding energy \cite{audi95} of $^{4}$He  and the Coulomb contributions  taken
from Ref. 22, where they  were determined from a 
least-squares fit to 
experimental data. 

In Table 1 we compare the experimental ground-state
binding energies~\cite{audi95} with the calculated ones for three different values of
$b_{\rm out}$, 1.75, 2.00, and 2.25 fm. In the last column, labeled LBSM,  we also report  
the results  obtained from the large-basis shell-model calculations
of Ref. 4. In this work, a complete $(0+2) \hbar \omega$ model space 
and an empirical effective interaction were used.\cite{wolt90} 

From Table 1 we see that the binding energies of the He isotopes 
and their corresponding mirror nuclei are well reproduced by using 
$b_{\rm out}=2.25$ fm. For these nuclei we overestimate the experimental
values by about 0.5 MeV, which implies that a little larger $b_{\rm out}$
would be sufficient to reproduce them. As regards the other nuclei
reported in Table 1, the value  2.0 fm  gives 
a good agreement between theory and experiment. 
In particular, for $^{6}$Li the experimental energy is overestimated by 
0.360 MeV, while for $^{7}$Li and $^{7}$Be it is underestimated by about 
the same value. As regards $^{8}$Li and $^{8}$B, the experimental binding
energies are reproduced by the theory within few tens of keV.
It should be noted that with the above choice of $b_{\rm out}$ the agreement
between experiment and theory is comparable to, and in some cases better than, 
that obtained in Ref. 4.

\begin{table}[t]
\caption{Experimental and calculated ground-state binding energies (MeV). 
See text for details.}
\begin{center}
\begin{tabular}{|c|c|c|c|c|c|}  
\hline
{ \raisebox{0pt}[13pt][7pt]  {$^{A}Z$} }   &
{ \raisebox{0pt}[13pt][7pt]  { Expt} } &
{ \raisebox{0pt}[13pt][7pt]  {TFSM(1.75)}} &
{ \raisebox{0pt}[13pt][7pt]  {TFSM(2.00)}} &
{ \raisebox{0pt}[13pt][7pt] {TFSM(2.25)}} & 
{ \raisebox{0pt}[13pt][7pt]  { LBSM}} \\
\hline
\raisebox{0pt}[13pt][0pt]{{ $^{6}$He}} & 
\raisebox{0pt}[13pt][0pt]{29.27} &
\raisebox{0pt}[13pt][0pt]{31.57} & 
\raisebox{0pt}[13pt][0pt]{30.59} & 
\raisebox{0pt}[13pt][0pt]{29.81} &  
\raisebox{0pt}[13pt][0pt]{29.74}  \\
{ $^{6}$Be} & 26.92 & 29.21 & 28.23 & 27.45 &  --- \\
&&&&&\\
{ $^{6}$Li} & 31.99 & 33.80 & 32.35 & 31.13 &   31.24  \\
&&&&&\\
{ $^{7}$He} & 28.82 &31.25 & 30.17 & 
29.32 & --- \\
{ $^{7}$B} & 24.72 & 26.99 & 25.91 & 
25.06 & --- \\
&&&&&\\
{ $^{7}$Li} & 39.24 & 42.02 & 38.86 & 36.30 &
  40.04\\
{ $^{7}$Be} & 37.60 & 40.36 & 37.20 & 34.64 &
  --- \\
&&&&&\\
{ $^{8}$He} & 31.41 & 35.61 & 33.57 &31.91 &  31.24 \\
&&&&&\\
{ $^{8}$Li} & 41.28 & 45.07 & 41.23 & 38.16 &   41.54 \\
\raisebox{0pt}[0pt][9pt]{{ $^{8}$B}} &
\raisebox{0pt}[0pt][9pt]{37.74} & 
\raisebox{0pt}[0pt][9pt]{41.51} &
\raisebox{0pt}[0pt][9pt]{37.67}& 
\raisebox{0pt}[0pt][9pt]{34.60}& 
\raisebox{0pt}[0pt][9pt]{---}\\
\hline
\end{tabular}
\end{center}
\end{table}

We have then calculated the spectra and the electromagnetic properties 
of the nuclei reported in Table 1 
using $b_{\rm out}=2.0$ and 2.25 fm for Li and He isotopes (and
their corresponding mirror nuclei), respectively. It is worth noting that we 
have not tried to adjust the  value of $b_{\rm out}$ for each nucleus.
The results obtained are in satisfactory agreement with the
experimental data and will be reported in  a forthcoming paper.\cite{pub00} Here
we shall only focus attention on $^{8}$Li. As regards the mirror
nucleus $^{8}$B, only two excited states have been observed, one of them
without spin-parity assignment. 

In Fig. 1 the experimental and calculated
spectra of $^{8}$Li are compared. To evidence the scope of the TFSM, we
also report the results we have obtained in a one-frequency shell-model
calculation (OFSM) \cite{kuo99} as well as those of Ref. 4. In the OFSM  
calculations the length parameter, which is the same for the core and
valence orbits, was fixed at 1.9 fm. This choice leads to a binding energy of
41.32 and 37.76 MeV for $^{8}$Li and $^{8}$B, respectively. 

\begin{figure}[t]
\epsfxsize=25pc 
\epsfbox{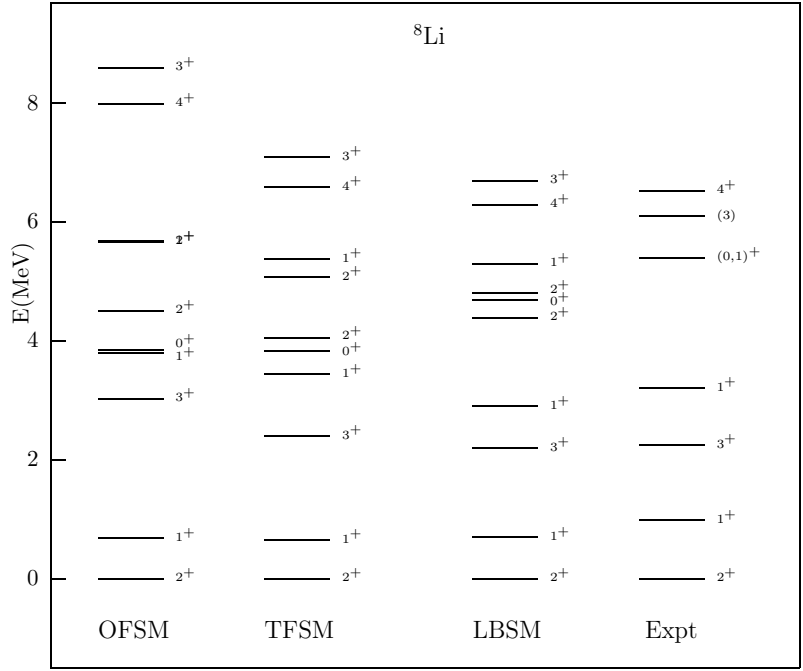} 
\caption{Experimental and calculated spectra  of $^{8}$Li. See text for details.
  \label{fig:radish}}
\end{figure}

From Fig. 1  we see that the three  calculations give rise to
the same levels, the OFSM and TFSM approaches  producing  also the 
same ordering. 
All the observed levels are predicted  by the different calculations.
They also yield, however,   
two more excited $2^+$ states  
 while for the $0^+$ and $1^{+}_{3}$ states only one
experimental level with spin equal to 0 or 1 is available. 
All three calculations  suggest that this state, which lies at 5.4 MeV, has $J^{\pi}=1^+$
while for  the (3) state at 6.1 MeV
they predict $J^{\pi}=3^{+}$. Regarding
the quantitative agreement, the excitation energies produced by the 
one-frequency calculations are significantly higher than the observed ones
for most states, the discrepancy between theory and experiment ranging 
from 0.3 to 2.5 MeV for the $1^{+}_{1}$ and the $3^{+}_{2}$, respectively.  
On the other hand, the agreement between the experimental
and the TFSM spectrum is very good. A significant difference (about 1 MeV)
occurs, in fact, only for the second $3^+$ state while  for all the other
states the discrepancies are less than 0.3 MeV.

As regards the quality of our results, Fig. 1 shows that they
are comparable to those of Ref. 4. This is quite an achievement if one
considers the very large model space used in the LBSM calculations.

\section{Summary}

In this paper, we have reported some results of  TFSM calculations for $p$-shell nuclei.
They have been obtained by  employing
an effective interaction derived from the Bonn-A nucleon-nucleon potential
by means of a $G$-matrix folded-diagram method. 

Within the framework of the TFSM, the $G$ matrix is calculated in a space composed of wave 
functions of two
different length parameters, one for the inner core orbits and the other for the outer valence
orbits. As  compared to large-basis shell-model calculations, this approach
provides  a convenient  alternative, as it appears from the results of our calculations.
The quality of these  results 
may be taken as an indication that most of the effects  which are not explicitly
taken into account in our model space are included in the
effective interaction.  
   
\section*{Acknowledgments}
This work was supported in part by the Italian Ministero dell'Universit\`a 
e della Ricerca Scientifica e Tecnologica
(MURST) and by the US grant no DE-FG02-88ER40388. N.I. would like to thank 
the European Social Fund for financial support.

\end{document}